\documentclass[epj]{webofc}
\usepackage[varg]{txfonts}   
\wocname{EPJ Web of Conferences}
\woctitle{CONF12}
\begin{document}
\selectlanguage{english}
\title{Flow anisotropy due to momentum deposition in ultra-relativistic nuclear collisions}
%
%

\author{Boris Tom\'a\v{s}ik\inst{1,2}\fnsep\thanks{\email{boris.tomasik@cern.ch}} \and
        Martin Schulc\inst{2,3} 
}

\institute{Univerzita Mateja Bela, Bansk\'a Bystrica, Slovakia
\and
           FNSPE, Czech Technical University, Prague 1, Czech Republic
\and
           Research Centre \v{R}e\v{z}, Husinec-\v{R}e\v{z}, Czech Republic
}

\abstract{%
Minijets and jets are produced in large numbers in nuclear collisions at TeV energies, 
so that there are many of them in a single fireball. They deposit non-negligible amount 
of momentum and energy into the hydrodynamically expanding bulk and cause anisotropies 
of the expansion. Moreover, due to their multiple production in a single event the resulting 
anisotropies are correlated with the collision geometry and thus contribute positively also 
to event-averaged anisotropies in non-central collisions. Using simulations with three-dimensional 
ideal hydrodynamic model we demonstrate the importance of this effect. It must be taken 
into account if conclusions about the properties of the hot matter are to be drawn.
}
\maketitle
\section{Introduction}
\label{intro}

One of the features of heavy-ion collisions at the LHC is the large portion of energy spent 
in production of hard partons. Some of them appear as jets, but a major part of hard and
semi-hard partons never comes out of the fireball. Instead, they are fully stopped in the 
quark-gluon plasma. The energy as well as momentum of the partons is thus fully transformed into 
the fluid medium. It has been shown that this generates streams in the plasma which continue 
to move even after the (originally) hard partons have thermalised \cite{betz}. 
It is reasonable to expect that such streams would contribute to flow anisotropy 
of the created quark matter. 

\begin{figure}[h]
\centering
\sidecaption
\includegraphics[width=0.57\textwidth,clip]{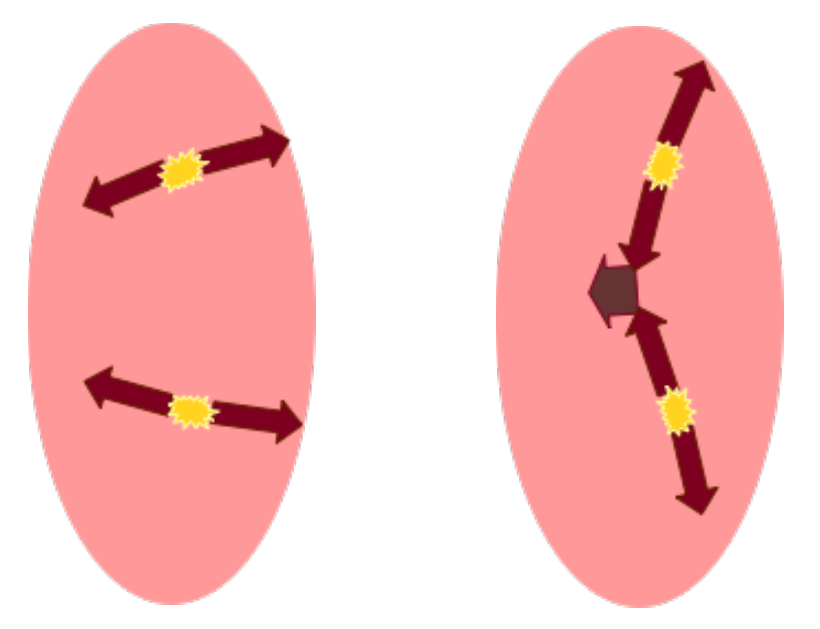}
\caption{Left: a pair of dijets directed in the reaction plane. Right: a pair of dijets oriented
out of the reaction plane.}
\label{f:cartoon}       
\end{figure}

There is more than just one pair of hard partons per event at the LHC. Therefore, we can also 
expect more streams within the expanding plasma. If their number is not too large, then we 
can expect an increase of flow anisotropies of all orders. In addition to this, we also argue that 
this contribution to elliptic flow anisotropy is correlated with the geometry of the collision
so that it actually increases the elliptic flow in non-central collisions. This can be explained 
with the help of the schematic drawings shown in Fig.~\ref{f:cartoon}.
%
%
In non-central collisions the elliptic flow is caused by larger pressure gradients in the 
direction of the reaction plane. They result in faster expansion in that direction.
If two dijets are both produced in the direction of the reaction plane, they both contribute 
positively to the elliptic flow.\footnote{For brevity, we will here refer to hard partons as to jets and 
back-to-back pairs of hard partons as dijets even if they have not yet formed the showers.}
Jets are produced anisotropically, however. That means, that there will also be a pair 
of dijets produced perpendicularly to the reaction plane. One would naively expect that 
they would then suppress the flow anisotropy. 
However, as it is sketched in Fig.~\ref{f:cartoon}, in this case the inward-flying jets are more likely 
to meet. Then the streams that they generate would merge, partially cancel each other and flow 
together in a new direction determined by their total momentum. Thus the resulting effect of the 
pair of dijets on the collective flow would be smaller than in the previous case, since a pair of dijets 
flying in the direction of the reaction plane is less likely to meet. 

We will show below with the help of hydrodynamic simulations that these handwaving arguments 
are indeed true. We shall also show that the effect of the jets on flow anisotropies is significant.

\section{The model}
\label{s:model}

We have constructed a three-dimensional ideal hydrodynamic model which includes a source term
in the energy-momentum conservation equation
\begin{equation}
\partial_\mu T^{\mu\nu} = J^\nu \,  .
\end{equation}
In fact, it might be better to call it a \emph{force} term, since this is the real meaning of that term. 

For the hydrodynamic evolution itself we chose the Equation of State \cite{EoS} which combines the 
results of Lattice QCD at high temperatures with the construction from a hadron resonance gas at low 
temperatures. To handle large gradients and shocks, the model uses the SHASTA algorithm
\cite{shasta}. 

The initial conditions that we have chosen are smooth, with the initial energy density profile specified 
by the optical Glauber model. Intentionally, we have chosen this obsolete solution. The reason is that 
we shall be implementing a novel mechanism which should lead to flow anisotropies. Not having 
fluctuations in the initial state makes it easier to evaluate its impact. The initial energy density in the 
central point of the fireball produced in collisions with vanishing impact parameter was 60~GeV/fm$^3$
and the time when hydrodynamics is started was set to $\tau_0 = 0.55$~fm/$c$.

The number of dijets fluctuates but on average we have about 10 pairs with $p_t$ above 3~GeV/$c$
in central Pb+Pb collision at $\sqrt{s_{NN}}=5.5$~TeV. The two jets of the pair are produced back-to-back
in $p_t$ but they generally have different rapidities. Their initial positions in transverse plane
are given by the distribution of the binary nucleon-nucleon collisions. Transverse momenta
of the jets follow the distribution \cite{levai}
\begin{equation}
\frac{1}{2\pi} \frac{d\sigma}{p_t\, dp_t\, dy} = \frac{B}{\left ( 1 + \frac{p_t}{p_0} \right )^n}
\end{equation}
with $B = 14.7$~mb/GeV, $p_0 = 6$~GeV, $n=9.5$.

The jets, as they traverse the quark-gluon plasma, lose their energy and momentum. It is assumed 
that this transfer scales with the entropy density of the medium
\begin{equation}
\frac{dE}{dx} = \left . \frac{dE}{dx}\right |_0 \frac{s}{s_0}
\end{equation}
where $s_0$ is the entropy density which corresponds to the energy density of 20~GeV/fm$^3$ and the 
energy loss scale $dE/dx|_0$ has been varied in order to investigate its influence on the observable results. 

The deposition of the energy and momentum has been smeared in space with the help of a Gaussian 
distribution with the width of 0.3~fm. We have checked that varying the width to 0.15~fm and 0.6~fm does 
not influence the results much. 

Cooper-Frye freeze-out was handled with the help of THERMINATOR2 package \cite{therm}. Production 
and decays of resonances are included.

\section{Results for non-central collisions}
\label{s:noncen}

\begin{figure}[h]
\centering
\sidecaption
\includegraphics[width=0.57\textwidth,clip]{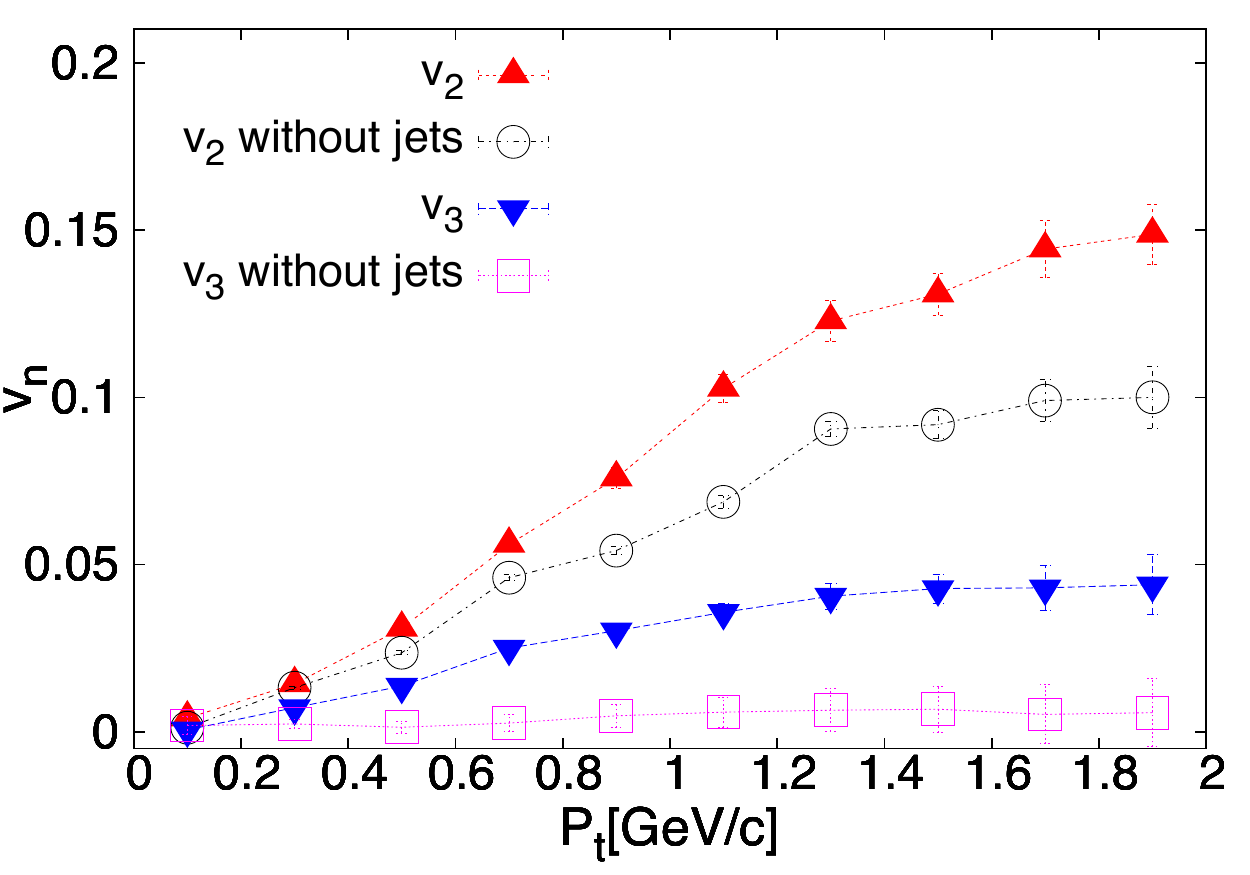}
\caption{Elliptic flow and triangular flow in Pb+Pb collisions, centrality class 20--30\%,
calculated with smooth fireball and no energy-momentum 
deposition from jets (circles and squares), as well as with energy-momentum deposition 
from jets (triangles).}
\label{f:noncen}       
\end{figure}

In order to confirm the hypothesis that the jets would enhance the elliptic flow in non-central 
collisions, we have performed simulations of Pb+Pb collisions within 30--40\% centrality class
\cite{jets}.
The main results are summarised in Fig.~\ref{f:noncen}.
%
%
There we plot the elliptic flow coefficient $v_2$ and the triangular flow coefficient $v_3$ as 
functions of $p_t$. They are both calculated for two scenarios. One is the simulation with 
jet energy loss included and the value of $dE/dx|_0$ set to 4~GeV/fm. For reference, we also 
show results from simulations with no jets. There, the elliptic flow results solely from the 
difference between the pressure gradients in-plane and out-of-plane in the initial 
conditions and there is no triangular flow. We see that the energy and momentum deposition 
enhances $v_2$ by about 50\%. This confirms our hypothesis: the flow anisotropy due to 
momentum deposition is correlated with the geometry of the collision via the effect described in the 
introductory section. Note that the jets also generate triangular anisotropy of the collective flow 
which otherwise would be absent in our simulations. 

The cartoon argumentation used in the introduction might also illustrate a possible way of 
distinguishing our mechanism of generation of the flow anisotropy from the more conventional 
scenario where any anisotropy is just due to anisotropies in energy density distribution in the 
initial state of the collision. Recall that the combined second and third-order anisotropy would 
be parametrised as 
\begin{equation}
\frac{dN}{d\phi} = \frac{N}{2\pi} \left (
1 + 2v_2 \cos\left ( 2(\phi -\psi_2)  \right )
+ 2v_3 \cos \left ( 3(\phi - \psi_3) \right )
\right )\,  .
\end{equation}
Figure~\ref{f:cartoon} suggests that the triangular flow is connected with merging of streams, 
which is more likely for the  jets flying out of the reaction plane. Then the third order reaction 
plane in an event is more likely to be directed together with the second order reaction plane.

\begin{figure}[h]
\centering
\sidecaption
\includegraphics[width=0.57\textwidth,clip]{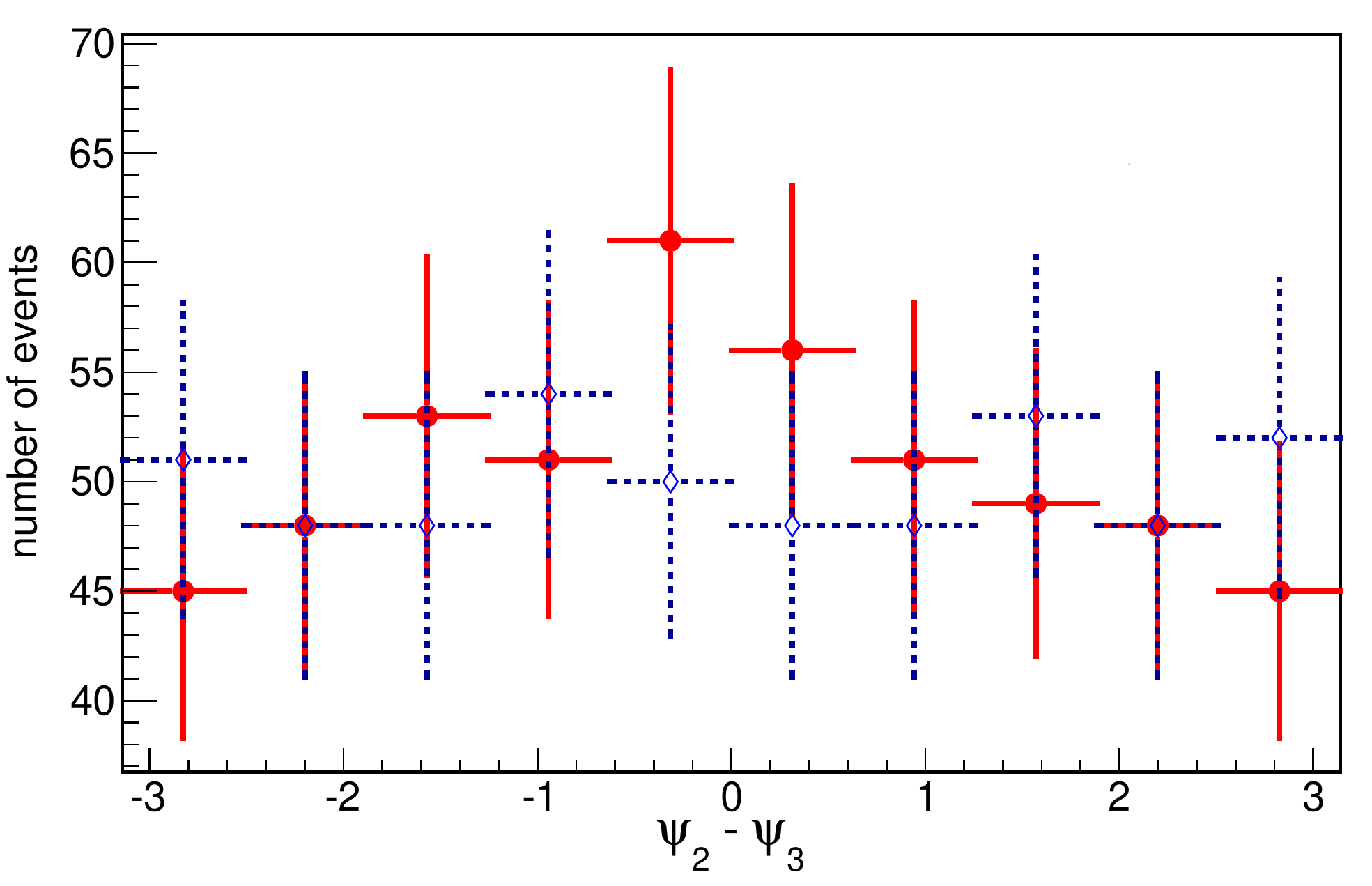}
\caption{Correlation function in the difference of second and third-order event planes. Determined 
for events with momentum deposition from jets (red solid circles) and for events with hot spots 
superimposed over smooth initial conditions (blue dashed open diamonds).}
\label{f:correl}       
\end{figure}

In order to test this idea we have generated also a set of events where instead of jets we have 
included anisotropies into the initial conditions only. The same amount of energy and momentum 
has been put within localised ``hot spots'' which were superimposed over the smooth 
energy density profile. We then measured the correlation function of the difference $\psi_2-\psi_3$,
which is plotted in Fig.~\ref{f:correl}.
%
%
Unfortunately, due to high computational cost we have only 500 events simulated for each 
of the two models, therefore the error bars are rather large and better statistics would be needed for a 
conclusive answer. However, the data are compatible with the hypothesis that momentum deposition from 
jets into plasma and the merging of streams lead to correlated $\psi_2$ and $\psi_3$, while 
such correlation is absent with the flow anisotropies caused only by initial conditions.

\section{Results for central collisions}
\label{s:central}

\begin{figure}[h]
\centering
\includegraphics[width=0.45\textwidth,clip]{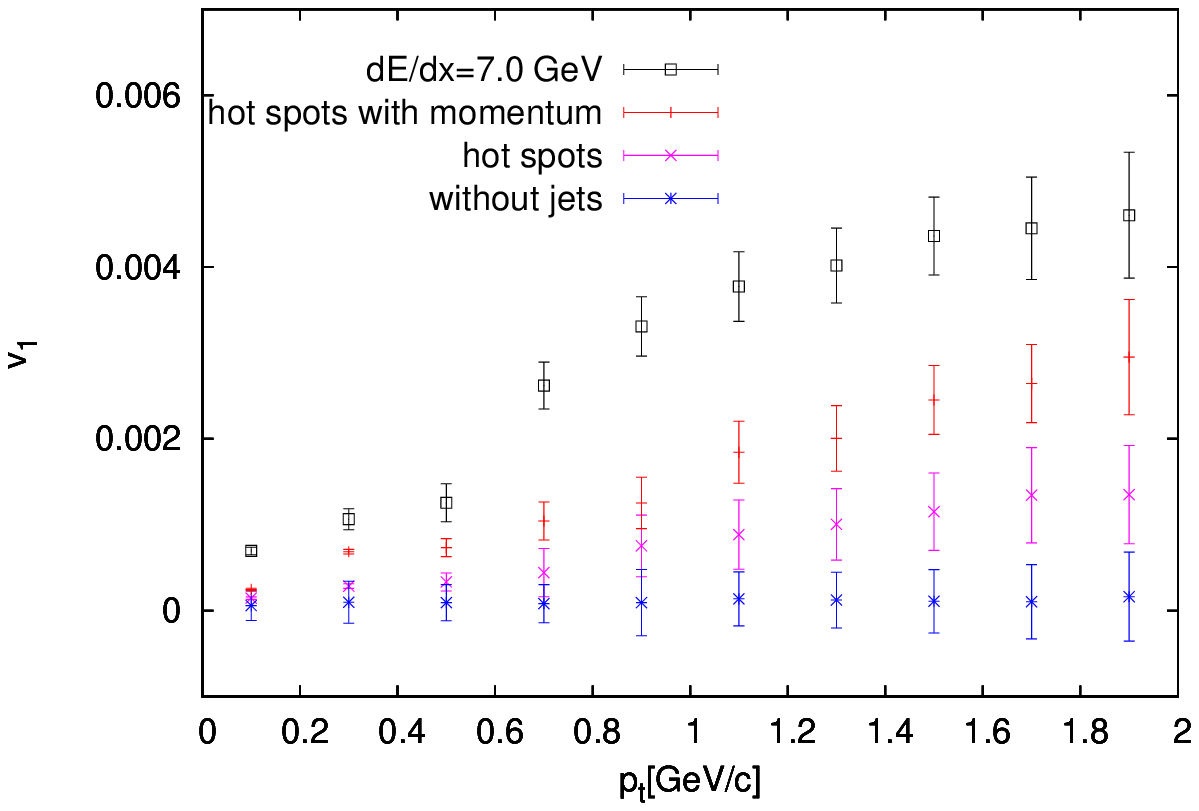}
\includegraphics[width=0.45\textwidth,clip]{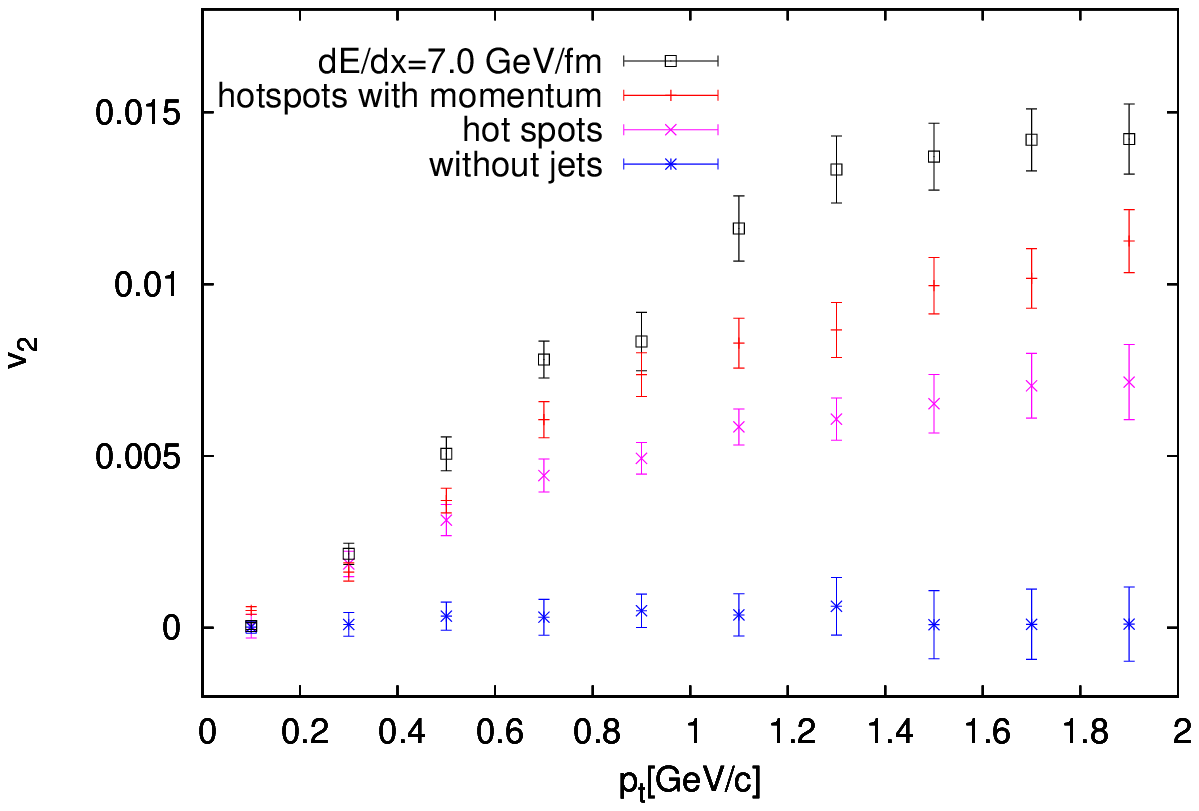}
\includegraphics[width=0.45\textwidth,clip]{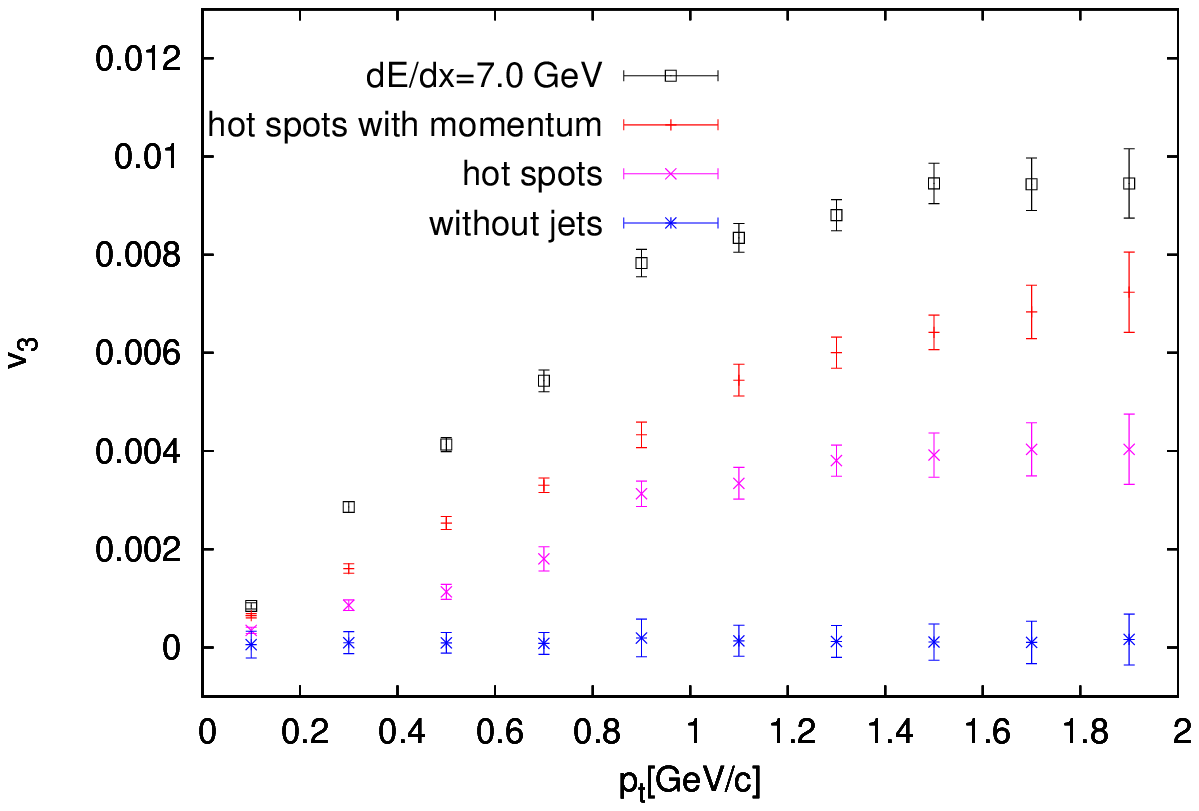}
\includegraphics[width=0.45\textwidth,clip]{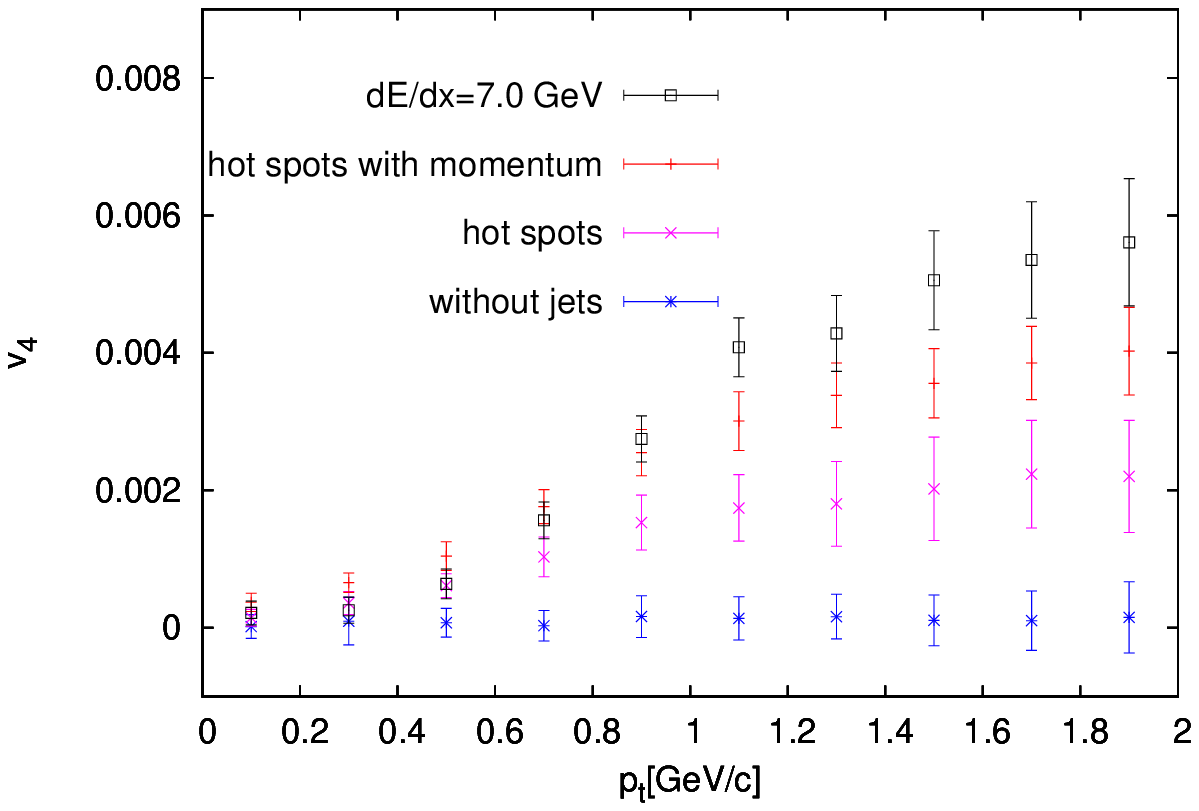}
\caption{Anisotropy parameters $v_1$, $v_2$, $v_3$, and $v_4$ calculated for Pb+Pb collisions
at $\sqrt{s_{NN}} = 5.5$~TeV at vanishing impact parameter. Four settings are compared:
only smooth initial conditions with no jets (blue stars), hotspots in the initial conditions and 
no jets (purple x's), as previous but the hotspots contain also momentum (red bars), energy and momentum 
deposition from jets (black squares).
}
\label{f:central}       
\end{figure}

Experimental data suggest that hadronic distributions show anisotropies in azimuthal angle even 
in ultra-central heavy-ion collisions. We investigated to what extent they can be generated by the 
mechanism proposed here. To this end, we simulated the evolution of fireballs which are created 
at vanishing impact parameter and let the jets lose energy and momentum there. The results are shown 
in Fig.~\ref{f:central} 
%

It is also interesting to explore whether or not the effect of jets can be mimicked just 
by including the corresponding anisotropies into the initial profile of the energy density. 
In order to answer this question we have  produced two additional sets of simulations. 
First, on top of the smooth energy density profile we superimposed places with increased
energy density, so that the energy of one such ``hot spot'' is equal to the energy of a dijet 
that would originate there. The results, also shown in Fig.~\ref{f:central}, are clear: this mechanism can 
account for not even 50\% of the anisotropies which are due to jets. One might argue, however, 
that in this simple hot spot scenario no \emph{momentum} anisotropies are initiated, whereas 
jets deposit also momentum into the fluid. Therefore, we have made another set of initial conditions
where the hot spots included not only additional energy, but also momentum. 
The included momentum was equal to the total momentum that would 
have been deposited by the jet. Nevertheless, neither this set of initial conditions did reproduce
the results from the simulations with jets. We conclude that momentum deposition \emph{during}
the evolution of the fireball cannot be mimicked by an augmented set of initial conditions
\cite{martin}.

\section{Conclusions}
\label{s:conc}

Momentum deposition from hard partons represents an important contribution to flow asymmetry in 
ultrarelativistic nuclear collisions. That asymmetry is often being used for the measurement of transport 
properties of quark-gluon plasma. In such a measurement, results of hydrodynamic simulations which depend on transport 
coefficients, are compared to data and the coefficients are tuned in order to reach an agreement. 
It is important to include jets in such calculations, if reliable quantitative results on the viscosities 
are to be obtained.


\paragraph{Acknowledgements}
We thank Ulrich Heinz for pointing out to us the possible correlation between $\psi_2$ and $\psi_3$
discussed in Fig.~\ref{f:correl}. 
This work has been partially supported by VEGA 1/0469/15 (Slovakia) and LG15001 (Czech Republic).

\end{document}